# Extreme multiexciton emission from deterministically assembled single emitter subwavelength plasmonic patch antennas

Amit Raj Dhawan[1,2], Cherif Belacel[2,5], Juan U. Esparza-Villa[2], Michel Nasilowski[3], Zhiming Wang[1], Catherine Schwob[2], Jean-Paul Hugonin[4], Laurent Coolen[2], Benoît Dubertret[3], Pascale Senellart[5], Agnès Maître[2*]

[1]Institute of Fundamental and Frontier Sciences, University of Electronic Science and Technology of China, Chengdu 610054, People's Republic of China

[2]Sorbonne Universités, UPMC Univ Paris 06, UMR 7588, Institut de NanoSciences de Paris (INSP), Paris F-75005, France

[3]Laboratoire de Physique et d'Etude des Matériaux, ESPCI-ParisTech, PSL Research University, Sorbonne Université UPMC Univ Paris 06, CNRS, 10 rue Vauquelin 75005 Paris, France

[4]Laboratoire Charles Fabry, Institut d'Optique Graduate School, CNRS UMR 8501, Université Paris-Sud, 2 avenue Augustin Fresnel, 91127 Palaiseau Cedex

[5] Centre de Nanosciences et de Nanotechnologies et de Nanostructures, CNRS UMR9001, Université Paris Sud, Université Paris-Saclay, Route de Nozay, 91460 Marcoussis, France.

**Abstract**

Plasmonic antennas are attractive optical structures for many applications in nano and quantum technologies. By providing enhanced interaction between a nanoemitter and light, they efficiently accelerate and direct spontaneous emission. One challenge, however, is the precise nanoscale positioning of the emitter in the structure. Here we present a laser etching protocol that deterministically positions a single colloidal CdSe/CdS core/shell quantum dot emitter inside a subwavelength plasmonic patch antenna with three-dimensional nanoscale control. By exploiting the properties of metal–insulator–metal structures at the nanoscale, the fabricated single emitter antenna exhibits an extremely high Purcell factor (>72) and brightness enhancement by a factor of 70. Due to the unprecedented quenching of Auger processes and the strong acceleration of multiexciton emission, more than 4 photons per pulse can be emitted by a single quantum dot. Our technology permits the fabrication of bright room-temperature single-emitter sources emitting either multiple or single photons.

**Introduction.** The interaction between an emitter and its local electromagnetic field can be engineered by increasing the local density of states for applications in quantum information [1] and single photon generation [2]. This has been widely explored with various dielectric or plasmonic environments [3, 4, 5, 6, 7] and a large variety of solid-state emitters such as self-assembled [8] or colloidal [9] quantum dots, single molecules [10], and defects in diamond [11]. Optimizing the coupling between the emitter and the nanostructure makes it possible to control the emission directivity, and the dynamics of spontaneous emission. The latter is quantified by the Purcell factor $F_p$, which scales as the inverse of the electromagnetic confinement volume provided by the photonic structure [12, 13, 4]. Plasmonic nano-antennas [14] exhibit very low volume and wide spectral resonance, and therefore are excellent structures for obtaining high Purcell factors $F_p$ with broadband emitters.

Achieving intense light–matter interaction using extremely low volume plasmonic structures [9, 15] necessitates very precise, hence challenging, spatial positioning of the emitter at the nanometer scale. Many works so far have relied on randomly positioned emitters to demonstrate high acceleration of spontaneous emission of a single quantum dot (QD) or a cluster of QDs coupled to a plasmonic antenna [16, 17, 18]. The challenge of deterministically positioning individual colloidal QDs inside nanophotonic structures with three-dimensional nanoscale control has to be overcome for realizing efficient devices working at room temperature and benefiting from highly optimized light-matter interaction regimes [19, 16].

A promising approach for the deterministic positioning of emitters in dielectric cavities or plasmonic antennas is in-situ optical lithography [20], which consists in measuring the emitter position optically through emission mapping with subwavelength accuracy and defining the photonic structure around the emitter during a single optical lithography step. It has been shown to allow the fabrication of efficient dielectric cavity based single photon source operating at cryogenic temperature [21], as well as to couple single self-assembled quantum dots [22] and small clusters of colloidal quantum dots in plasmonic nanostructures [6]. A similar approach has been developed using in-situ e-beam lithography [23] to create self-assembled QD microlens structures. Thus far, these technologies have not been transferred to chemically synthetized emitters such as colloidal QDs that have strong potential for cheap and room-temperature applications. The main reason for this is the exacerbated sensitivity of these emitters to technological processing. Because of the strong sensitivity of the emission process to surface states, preserving their properties during the technological protocols, involving, for example, e-beam or laser exposure, dry or wet etchings and solvents, is highly challenging.

Here, we report on a non-destructive in-situ far-field laser etching lithography technique that allows controlling—at the nanometer scale—the position of fragile emitters in plasmonic structures. Our technique makes use of multilayer structures and a dual wavelength protocol to deterministically position a single CdSe/CdS colloidal QD in a plasmonic patch antenna and fabricate subwavelength antennas of different shapes and sizes achieving state-of-the-art Purcell factor on a single emitter. Strong electromagnetic interactions in the plasmonic antenna decreases the relative efficiency of Auger recombination channels, which results in very high brightness and extreme multiexciton radiative recombination. The technique can be used to fabricate a variety of photonic structures embedding fragile emitters.

**Plasmonic patch antenna: design and operation.** Figure 1(a) depicts the system that we explore in this work: a single emitter coupled to a plasmonic patch antenna. The antenna consists of a thin dielectric layer (typically 30–40 nm) that is sandwiched between an optically thick bottom layer of gold, and a thin gold patch on top; the patch thickness is about 20 nm and its diameter is in a range of 0.2–2.5 µm. This system has been both theoretically and experimentally shown to be an excellent tool for accelerating and directing fluorescence emission [24, 25].

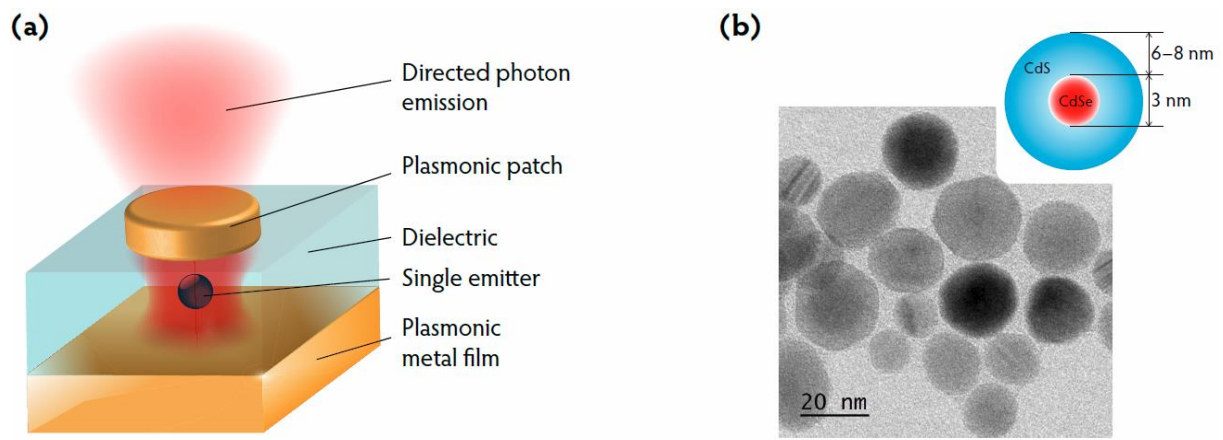

*Figure 1 | Plasmonic patch antenna coupled to a single colloidal quantum dot (a) Schematic of the structure under study consisting of a single colloidal core/shell QD coupled to a plasmonic patch antenna. (b) Schematic and transmission electron microscope image of the investigated emitters.*

Optimal positioning of the emitter inside the antenna couples its radiation to surface plasmon polaritons (SPPs) at both nano-spaced metal-dielectric interfaces. These SPPs at the two interfaces further couple and create strong confinement of the electromagnetic field around the emitter. The SPPs generated in the thin plasmonic metal patch (thinner than the skin depth) lead to the emission of photons [24] as depicted in Figure 1(a). The antenna operation depends on the dipole orientation of the emitter (stronger acceleration of spontaneous emission for vertical dipole orientation), the patch size (large patches are more directive but small patches show stronger resonances), and the dielectric spacer [24]. In this work, we insert chemically synthetized [26] relatively large quasi-spherical individual CdSe/CdS semiconductor core/shell colloidal QDs [27, 28] in the antenna. They have CdSe cores of about 3 nm in diameter, which are encapsulated by slowly grown 6–8 nm thick CdS shells (Figure 1(b)), which make them almost non-blinking [29]. Due to their high absorption cross-section, they show bright fluorescence at room temperature. Under ultra-violet (UV) excitation at room-temperature, they emit at 633 nm with a spectral width of 30 nm. We characterize these QDs in the low excitation limit, where Auger processes are efficient enough to lead to single photon emission[1].

**In-situ subwavelength laser etching on fragile emitters.** The proposed technological protocol is illustrated in Figure 2. It consists of deterministic and non-destructive in-situ laser etching that allows positioning a single QD within an antenna with a 3 nm vertical and 50 nm lateral precision (see Supplementary Information). We use a low luminescence bi-layer polymer [30] to be able to locate single emitters by mapping their luminescence.

The process starts by evaporating a thick layer of gold (200 nm) on a Si substrate using an intermediate adhesion layer of Ti/Cr. A thin layer of PMMA (10 nm) is then spin-coated above it. Then a layer of spatially well-isolated single QDs is spin-coated using a dispersion of QDs in hexane. To create optimized spacing around the single emitters, and to protect them from direct dielectric vapor deposition, a 35 nm smooth PMMA film is spin-coated, which embeds them in a dielectric layer. This protects the emitters in the proceeding lithographic steps. A bi-layer consisting of a lift-off resist (LOR) and PMMA is then spin-coated. The low luminosity of this bi-layer permits the detection of the emission of single QDs embedded beneath and excited by a low intensity spectrally filtered broadband supercontinuum laser emitting at 473–478 nm (Figure 2(a)). As in in-situ lithography methods [20, 23], we map QD fluorescence with nanoscale accuracy. The laser wavelength is then tuned to 550–605 nm, which corresponds to a wavelength range where the laser light is absorbed substantially more by the lift-off resist than by the QD. The excitation power can thus be chosen to burn the resist bi-layer locally and create a hole directly above the selected QD without photobleaching it (Figure 2(b)). We note in Figure 2(f) that the emission lifetime of a QD does not change after the resist burning—this demonstrates that the laser etching process does not photodegrade the QD. After chemical development (Figure 2(c)), gold is deposited by evaporation (Figure 2(d)), and a lift-off is performed to obtain patch antennas embedding a single emitter (Figure 2(e)). The atomic force microscopy (AFM) image above Figure 2(e) shows an antenna patch after the final lift-off.

---

[1] We note that under strong pumping, these large single QDs emit multiphotons because their multiexciton emission rate becomes comparable with the non-radiative Auger rate (see Supplementary Information).

This laser etching technique presents several advantages as compared to previous methods. A standard optical in-situ lithography would require intense UV laser light for exposure that degrades the QDs due to excessive light absorption. Here, by using a selected 550–605 nm wavelength for the laser etching, we avoid such detrimental effect. Moreover, this technique allows scanning and locating the emitter without exposing the photoresist. Finally, as the etching process is highly nonlinear with the excitation power, subwavelength structures can be obtained. The AFM image and the height profile of the antenna patch of Figures 3(a) and 3(b) show a nano-antenna with dimensions as small as 200×450 nm$^2$ while the mean wavelength of the etching laser was 580 nm.

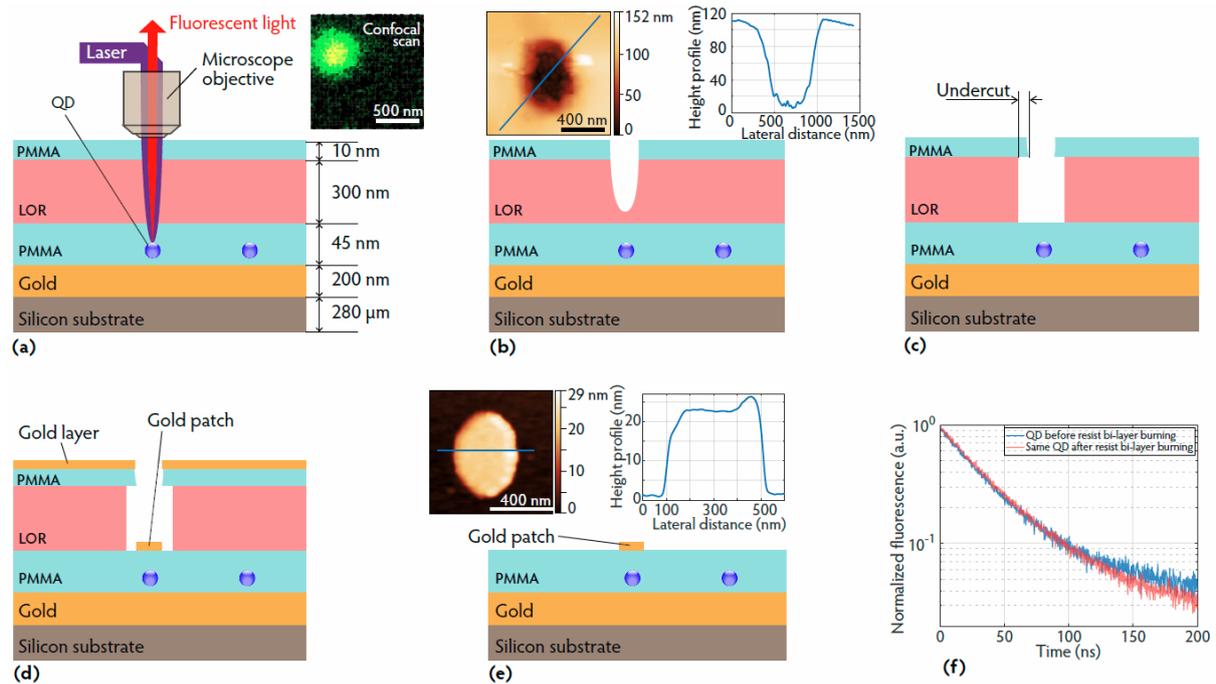

*Figure 2 | In-situ laser etching for deterministic assembly of single emitter plasmonic antenna nanostructures. Figures (a) to (e) illustrate the in-situ laser etching protocol used to fabricate the antennas (see details in the text). The image at the top right of (a) shows the confocal image of emission from a QD, and the image on top of (b) shows the AFM topography and height profile of a hole etched by the laser into the resist bi-layer. The undercut created by the selective etching of LOR (c) allows the lift-off at step (d) of lithography to obtain single emitter antennas (e). Above (e) are the AFM image and the corresponding height profile of a fabricated antenna patch. (f) Emission lifetime of a QD before (blue) and after (red) the laser etching.*

**Emission from a subwavelength size antenna.** We study the emission pattern of a very small antenna (Figures 3(a) and 3(b)) coupled to a single QD. Such subwavelength plasmonic patch antennas have been predicted to show very strong Purcell effect, and directive emission [24].

Figures 3(c) and 3(d) show the radiation pattern of the antenna measured in far-field by Fourier plane imaging. The antenna radiates through a cone in the far-field, centered around 45°. This measurement is in good agreement with numerical simulation of an elliptical patch antenna using the Fourier modal method [31, 32, 33, 34] as depicted in Figures 3(e) and 3(f) for the resonance wavelength at 604 nm.

A striking feature of the measured radiative pattern is its high symmetry in the far-field. Indeed, if the emitter is slightly off-centered with respect to the antenna patch, the radiation pattern rapidly shows angular asymmetry [6]. The symmetry in the lobes of Figure 3(d) and 3(f) demonstrates that the emitter is centered with respect to the antenna within 5 nm, thus exhibiting the precision of the fabrication method performed to position the QD laterally at the center of the patch.

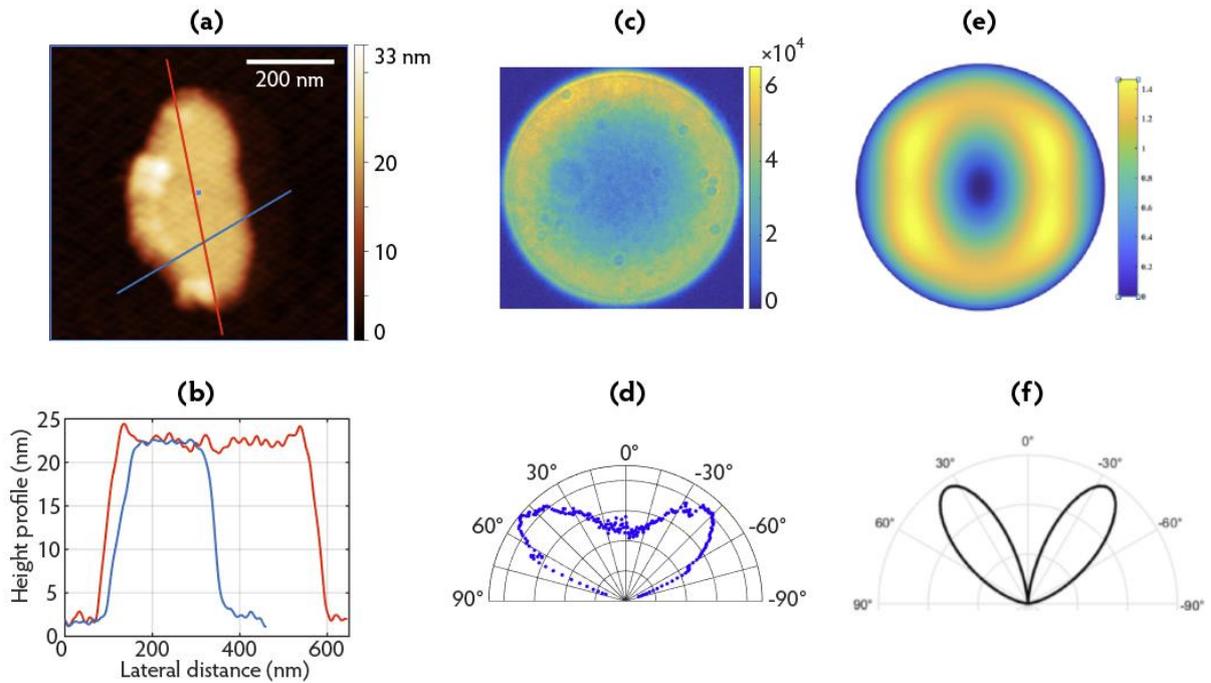

*Figure 3 | Topography and emission pattern of the emitter-antenna structures. (a) AFM image of a small patch antenna and (b) corresponding height profiles. (c) EMCCD (electron-multiplied charge-coupled device) camera image of the radiation pattern of the same antenna measured in far-field by imaging the Fourier plane of the QD emission, and (d) associated polar plot. (e) (θ,φ) polar plot of the numerically simulated emission pattern of a similar elliptical patch antenna and (f) polar plot along the maximum intensity axis.*

**Acceleration of spontaneous emission.** Figure 4(a) displays the fluorescence decay of a given QD before and after it was placed inside the patch antenna of Figure 3. Before the deposition of the upper gold patch, when the QD is embedded in a PMMA layer and is 10 nm above the gold surface, its emission decay has monoexponential character, that is indicative of single exciton recombination with a decay of $\tau_X^{\text{ref}} = 36$ ns. As proximity to gold shortens the emitter lifetime [35, 36, 37, 38], we estimate that, on average, this lifetime is 3 times shorter than the decay time of the same colloidal QDs in a homogeneous PMMA layer, i.e., $\frac{\tau_X^{\text{homogeneous}}}{\tau_X^{\text{ref}}} = 3$ (see Supplementary Information), where $\tau_X^{\text{homogeneous}}$ and $\tau_X^{\text{ref}}$ are the exciton lifetimes of the emitter in homogeneous medium (infinite dielectric medium of index 1.5) and reference medium (inside the same dieletric media but 10 nm above gold). After the deposition of a gold patch centered on the QD, the emission lifetime is considerably reduced to $\tau_X^{\text{Antenna}}$. The Purcell factor is calculated using the exciton decay rate, and is given as $F_P = \frac{\tau_X^{\text{homogeneous}}}{\tau_X^{\text{ref}}} \times \frac{\tau_X^{\text{ref}}}{\tau_X^{\text{Antenna}}}$.

Note that the acceleration of spontaneous emission observed in Figure 4(a) is so strong that we are limited by the response time of the measuring instrument from precisely quantifying $F_p$. By analyzing the instrument response function, we find that emission acceleration is greater than 24 with respect to the lifetime of the same QD before the gold patch deposition, which results in $F_p$ > 72 (see Supplementary Information), when comparing with the QD embedded in a homogeneous dielectric medium. We remark that the stated $F_p$ was found by comparing the exciton lifetime of the QD with the slowest component (1.5 ns) of the instrument response function and not the faster component (0.3 ns)—this results in a lower but more accurate $F_p$. The method is detailed and justified in the Supplementary Information. In comparison to former works [6, 9, 39], this extreme Purcell effect for a single emitter constitutes a new state-of-the-art value that is obtained by the controlled and deterministic positioning of the emitter at nanometric scale in a highly confining subwavelength antenna structure.

**Quenching of Auger processes and multiexciton emission.** As the consequence of such a significant Purcell effect, the electromagnetic decay channels are accelerated, making radiative multiexciton recombination more efficient than the Auger non-radiative channels. The photon correlation curves of Figure 4(b) shows the measured second order intensity correlation of the QD emission before it was placed in the antenna, exhibiting single photon emission with $g^{(2)}(0)$=0.2. When inserted in the antenna, this value rises to $g^{(2)}(0)$=1 (Figure 4(c)), confirming the radiative multiexciton recombination.

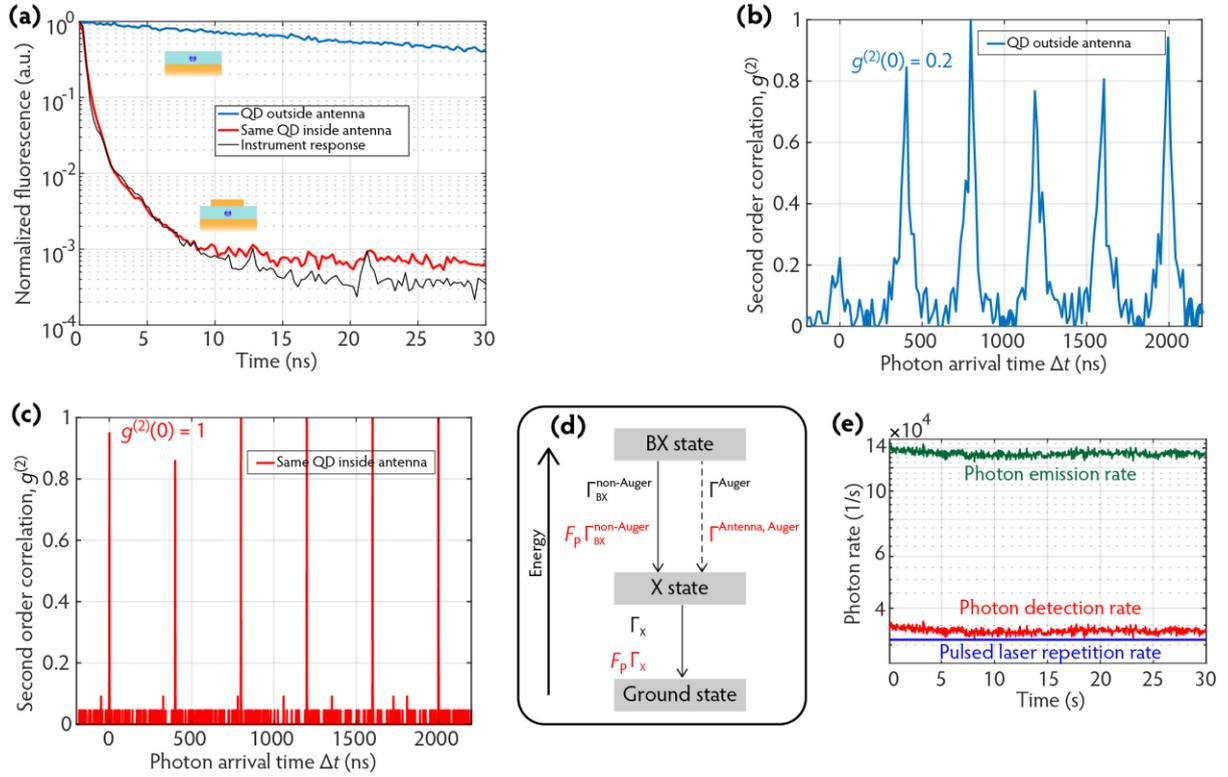

***Figure 4 | Emission characteristics of a highly accelerated antenna. (a)** Fluorescence decay of the QD before (blue line) and after (red line) it was placed inside the antenna. The instrument response function is shown in black. **(b)** Second order photon correlation measured on the QD before insertion in the antenna, and **(c)** inside the antenna under pulsed excitation of 0.03 W/cm² at 405 nm. **(d)** QD energy-level diagram showing biexciton (BX), exciton (X), and ground states. **(e)** Detected photon rate (red) and corresponding photon emission rate after detection efficiency correction (green). The blue line denotes the pulsed laser repetition rate.*

Inside the plasmonic antenna, the radiative decay rates are accelerated by the Purcell factor $F_P$, not only at the exciton but at the multiexciton level as well. As a result, Auger processes that are very efficient at the multiexciton level become relatively inefficient in the high Purcell factor regime, which facilitates radiative recombination of multiexcitons [40, 41, 42]. This is confirmed by the loss of single photon emission in Figure 4(c). The plot of Figure 4(e) illustrates very strong radiative multiexciton emission from a single emitter plasmonic optical antenna. Under 405 nm pulsed excitation at a rate of 31.25 kHz, the photon rate detected by the photodiodes is about 34 kHz. Given the 25% detection efficiency of our experimental setup, the photon emission rate from the QD is estimated to be about 136 kHz. Therefore, for each laser pulse, the single emitter in the antenna emits about 4–5 photons. This demonstrates extremely efficient quenching of Auger processes for at least 5 levels of radiative multiexciton recombination. It is very probable that the relaxation of the QD inside the antenna included many more higher multiexciton levels for which the non-radiative Auger recombination was not completely quenched, and which therefore did not contribute to radiation.

While high Purcell effect in plasmonic antennas has often been associated to low brightness due to the predominance of non-radiative channels [43, 44, 45, 46], the structures developed here display very high fluorescence enhancement. Figure 5(a) and (b) are CCD (charge-coupled device) camera fluorescence images of the same area including the small antenna of Figure 3 and some QDs scattered outside the antenna. Under lamp excitation (438±12 nm), in 10 ms of acquisition time, only the antenna emission (indicated by the yellow circle in Figure 5(a)) can be detected. An acquisition time of 200 ms (Figure 5(b)) is required to detect emission from QDs outside antennas. To quantify the brightness of the antenna, we use a pulsed laser at 405 nm to excite the QD before and after insertion in the antenna. Under the same excitation power, the antenna signal is 72 times more intense (Figure 5(c)).

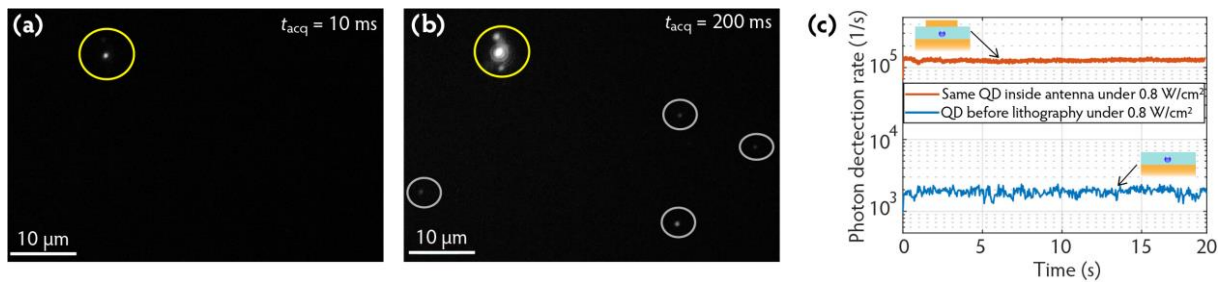

*Figure 5 | Fluorescence enhancement by a plasmonic antenna. (a, b) Fluorescence microscopy image of the antenna excited by a mercury lamp at 438±12 nm and captured by a CCD camera with an acquisition time of 10 ms (a) or 200 ms (b). The fluorescing antenna is encircled in yellow. Note that the two lighter spots above and below the saturated antenna spot in (b) are reflections on protecting glass before the CCD camera sensor. (c) Photon detection rate from the QD before (blue) and after (red) it was placed inside the antenna. The antenna is excited at 405 nm with a pulsed laser at 2.5 MHz.*

We also note the high optical quality and stability of the fabricated antenna. At this excitation, the antenna could sustain pulses with an instantaneous power of 120 mW for more than 10 minutes before it photobleached—exhibiting the quality of the QDs under study and their preservation during the laser etching protocol reported here. We finally note that the onset of multiexciton emission depends strongly on the antenna size and the Purcell effect it provides. In general, the appearance of multiexciton emission depends on the relative weight of the radiative decay of the multiexciton state and the Auger rate. When increasing the number of excitons, the corresponding multiexciton emission rate increases. In the presence of efficient Auger processes, it is thus possible to observe multiexciton emission when increasing the excitation power to the extent that the multiexciton radiative decay rate exceeds the Auger rate. When shortening the lifetime through the Purcell effect, the number of excitons needed to reach this regime decreases. This is what we observe studying several antennas of different Purcell factors. For larger (hence slower) antennas, the onset of multiexciton emission is typically observed at larger excitation power than for smaller (hence faster) antennas. This effect can be used in applications that demand single photon to multiphoton emission switching [47].

**Conclusion.** We have developed a novel deterministic laser etching protocol that allows the fabrication of subwavelength plasmonic antennas centered with 50 nm lateral accuracy on a single colloidal QD. The strong plasmonic coupling realized by controlled fabrication results in very bright and fast nanoantennas benefitting from multiexciton emission and directivity—this effect can be used to create efficient luminescent solar concentrators [48, 49], which can improve solar device efficiency. Thanks to the low modal volume and high field confinement, such an antenna shows very high Purcell effect and brightness, and directive emission. The high Purcell effect can lead to very pronounced multiexciton emission. Further developments of the protocol include selection of well-oriented emitters by polarization analysis [50, 51, 52] for achieving fully deterministic matching in spectrum, location and orientation between the emitter and the antenna modes for optimized light–matter interactions.

 The same protocol can be used to synthesize other nanophotonic devices which require deterministic and controlled positioning for a variety of sensitive emitters, working at room-temperature, such as small colloidal QDs (CdSe/CdS, PbSe/PbS, etc.), nanodiamonds with nitrogen-vacancy centers, and fluorescent molecules. Very bright and directive single photon sources based on antennas can be obtained with the same technique by custom designing the structure for a reduced Purcell factor or by using smaller colloidal QDs showing enhanced sensitivity to Auger processes.

**References**

[1]     M. S. Tame, K. McEnery, S. Özdemir, J. Lee, S. Maier, and M. Kim, "Quantum plasmonics," *Nature Physics*, vol. 9, no. 6, pp. 329–340, 2013.

[2]     I. Aharonovich, D. Englund, and M. Toth, "Solid-state single-photon emitters," *Nature Photonics*, vol. 10, no. 10, p. 631, 2016.

[3]     J. M. Gérard, B. Sermage, B. Gayral, B. Legrand, E. Costard, and V. Thierry-Mieg, "Enhanced spontaneous emission by quantum boxes in a monolithic optical microcavity," *Phys. Rev. Lett.*, vol. 81, pp. 1110–1113, Aug 1998. [Online]. Available: https://link.aps.org/doi/-10.1103/PhysRevLett.81.1110

[4]     G. Solomon, M. Pelton, and Y. Yamamoto, "Modification of spontaneous emission of a single quantum dot," *physica status solidi (a)*, vol. 178, no. 1, pp. 341–344, 2000.

[5]     P. Lodahl, A. F. Van Driel, I. S. Nikolaev, A. Irman *et al.*, "Controlling the dynamics of spontaneous emission from quantum dots by photonic crystals," *Nature*, vol. 430, no. 7000, p. 654, 2004.

[6]     C. Belacel, B. Habert, F. Bigourdan, F. Marquier, J.-P. Hugonin, S. M. de Vasconcellos, X. Lafosse, L. Coolen, C. Schwob, C. Javaux, B. Dubertret, J.-J. Greffet, P. Senellart, and A. Maitre, "Controlling spontaneous emission with plasmonic optical patch antennas," *Nano Letters*, vol. 13, no. 4, pp. 1516–1521, 2013, pMID: 23461679. [Online]. Available: http://dx.doi.org/-10.1021/nl3046602

[7]     F. Bernal Arango, A. Kwadrin, and A. F. Koenderink, "Plasmonic antennas hybridized with dielectric waveguides," *ACS Nano*, vol. 6, no. 11, pp. 10156–10167, 2012, pMID: 23066710. [Online]. Available: https://doi.org/10.1021/nn303907r

[8]     P. Senellart, G. Solomon, and A. White, "High-performance semiconductor quantum-dot single-photon sources," *Nature nanotechnology*, vol. 12, no. 11, p. 1026, 2017.

[9]     B. Ji, E. Giovanelli, B. Habert, P. Spinicelli, M. Nasilowski, X. Xu, N. Lequeux, J.-P. Hugonin, F. Marquier, J.-J. Greffet, and B. Dubertret, "Non-blinking quantum dot with a plasmonic nanoshell resonator," *Nat Nano*, vol. 10, no. 2, pp. 170–175, Feb 2015, article. [Online]. Available: http://dx.doi.org/10.1038/nnano.2014.298

[10]    R. Chikkaraddy, B. de Nijs, F. Benz, S. J. Barrow, O. A. Scherman, E. Rosta, A. Demetriadou, P. Fox, O. Hess, and J. J. Baumberg, "Single-molecule strong coupling at room temperature in plasmonic nanocavities," *Nature*, vol. 535, no. 7610, p. 127, 2016.

[11]    I. Aharonovich, A. D. Greentree, and S. Prawer, "Diamond photonics," *Nature Photonics*, vol. 5, no. 7, p. 397, 2011.

[12]    E. M. Purcell, "Spontaneous emission probabilities at radio frequencies," vol. 69, 1946, pp. 681+. [Online]. Available: http://prola.aps.org/pdf/PR/v69/i11-12/p674_2

[13]    G. Grynberg, A. Aspect, and C. Fabre, *Introduction to Quantum Optics: From the Semi-classical Approach to Quantized Light*, ser. Introduction to Quantum Optics: From the Semi-classical Approach to Quantized Light. Cambridge University Press, 2010.

[14]    A. Kinkhabwala, Z. Yu, S. Fan, Y. Avlasevich, K. Müllen, and W. Moerner, "Large single-molecule fluorescence enhancements produced by a bowtie nanoantenna," *Nature Photonics*, vol. 3, no. 11, pp. 654–657, 2009.

[15]    E. B. Ureña, M. P. Kreuzer, S. Itzhakov, H. Rigneault, R. Quidant, D. Oron, and J. Wenger, "Excitation enhancement of a quantum dot coupled to a plasmonic antenna," *Advanced Materials*, vol. 24, no. 44, pp. OP314–OP320, 2012. [Online]. Available: http://dx.doi.org/-10.1002/adma.201202783

[16]    T. B. Hoang, G. M. Akselrod, and M. H. Mikkelsen, "Ultrafast room-temperature single photon emission from quantum dots coupled to plasmonic nanocavities," *Nano Letters*, vol. 16, no. 1, pp. 270–275, 2016, pMID: 26606001. [Online]. Available: http://dx.doi.org/10.1021/-acs.nanolett.5b03724

[17]    Q. Le-Van, X. Le Roux, T. V. Teperik, B. Habert, F. Marquier, J.-J. Greffet, and A. Degiron, "Temperature dependence of quantum dot fluorescence assisted by plasmonic nanoantennas," *Phys. Rev. B*, vol. 91, p. 085412, Feb 2015. [Online]. Available: https://link.aps.org/doi/-10.1103/PhysRevB.91.085412

[18]    T. B. Hoang, G. M. Akselrod, and M. H. Mikkelsen, "Ultrafast spontaneous emission source using plasmonic nanoantennas," *Nature communications*, vol. 6, 2015.

[19]    K. J. Vahala, "Optical microcavities," *Nature*, vol. 424, no. 6950, pp. 839–846, Aug. 2003. [Online]. Available: http://dx.doi.org/10.1038/nature01939

[20]    A. Dousse, L. Lanco, J. Suffczynski, E. Semenova, A. Miard, A. Lematre, I. Sagnes, C. Roblin, J. Bloch, and P. Senellart, "Controlled light-matter coupling for a single quantum dot embedded in a pillar microcavity using far-field optical lithography," *Phys. Rev. Lett.*, vol. 101, p. 267404, Dec 2008. [Online]. Available: https://link.aps.org/doi/10.1103/PhysRevLett.101.267404

[21]    N. Somaschi, V. Giesz, L. De Santis, J. Loredo, M. P. Almeida, G. Hornecker, S. L. Portalupi, T. Grange, C. Antón, J. Demory *et al.*, "Near-optimal single-photon sources in the solid state," *Nature Photonics*, vol. 10, no. 5, p. 340, 2016.

[22]    O. Gazzano, S. M. De Vasconcellos, K. Gauthron, C. Symonds, J. Bloch, P. Voisin, J. Bellessa, A. Lematre, and P. Senellart, "Evidence for confined Tamm plasmon modes under metallic

microdisks and application to the control of spontaneous optical emission," *Physical review letters*, vol. 107, no. 24, p. 247402, 2011.

[23]     M. Gschrey, A. Thoma, P. Schnauber, M. Seifried, R. Schmidt, B. Wohlfeil, L. Krüger, J.-H. Schulze, T. Heindel, S. Burger *et al.*, "Highly indistinguishable photons from deterministic quantum-dot microlenses utilizing three-dimensional in situ electron-beam lithography," *Nature communications*, vol. 6, p. 7662, 2015.

[24]     R. Esteban, T. V. Teperik, and J.-J. Greffet, "Optical patch antennas for single photon emission using surface plasmon resonances," *Phys. Rev. Lett.*, vol. 104, p. 026802, Jan 2010. [Online]. Available: http://link.aps.org/doi/10.1103/PhysRevLett.104.026802

[25]     L. Novotny and N. van Hulst, "Antennas for light," *Nature Photonics*, vol. 5, no. 2, pp. 83–90, 2011. [Online]. Available: http://www.nature.com/nphoton/journal/v5/n2/full/­nphoton.2010.237.html

[26]     X. Peng, J. Wickham, and A. P. Alivisatos, "Kinetics of II-VI and III-V Colloidal Semiconductor Nanocrystal Growth: "Focusing" of Size Distributions," *Journal of the American Chemical Society*, vol. 120, no. 21, pp. 5343–5344, Jun. 1998. [Online]. Available: http://­dx.doi.org/10.1021/ja9805425

[27]     P. Reiss, M. Protière, and L. Li, "Core/shell semiconductor nanocrystals," *Small*, vol. 5, no. 2, pp. 154–168, 2009. [Online]. Available: http://dx.doi.org/10.1002/smll.200800841

[28]     V. Klimov, *Nanocrystal Quantum Dots*, 2nd ed. CRC Press, 2010.

[29]     B. Mahler, P. Spinicelli, S. Buil, X. Quelin, J.-P. Hermier, and B. Dubertret, "Towards non-blinking colloidal quantum dots," *Nature Materials*, vol. 7, no. 8, pp. 659–664, Jun. 2008. [Online]. Available: http://dx.doi.org/10.1038/nmat2222

[30]     E. Forsén, P. Carlberg, L. Montelius, and A. Boisen, "Laser lithography on resist bi-layer for nanoelectromechanical systems prototyping," *Microelectron. Eng.*, vol. 73–74, no. 1, pp. 491–495, Jun. 2004. [Online]. Available: http://dx.doi.org/10.1016/j.mee.2004.03.023

[31]     J. Yang, J.-P. Hugonin, and P. Lalanne, "Near-to-far field transformations for radiative and guided waves," *ACS Photonics*, vol. 3, no. 3, pp. 395–402, 2016. [Online]. Available: http://­dx.doi.org/10.1021/acsphotonics.5b00559

[32]     J. P. Hugonin and P. Lalanne, "Perfectly matched layers as nonlinear coordinate transforms: a generalized formalization," *J. Opt. Soc. Am. A*, vol. 22, no. 9, pp. 1844–1849, Sep 2005. [Online]. Available: http://josaa.osa.org/abstract.cfm?URI=josaa-22-9-1844

[33] Q. Bai, M. Perrin, C. Sauvan, J.-P. Hugonin, and P. Lalanne, "Efficient and intuitive method for the analysis of light scattering by a resonant nanostructure," *Opt. Express*, vol. 21, no. 22, pp. 27371–27382, Nov 2013. [Online]. Available: http://www.opticsexpress.org/abstract.cfm?URI=oe-21-22-27371

[34] A. David, H. Benisty, and C. Weisbuch, "Fast factorization rule and plane-wave expansion method for two-dimensional photonic crystals with arbitrary hole-shape," *Phys. Rev. B*, vol. 73, p. 075107, Feb 2006. [Online]. Available: https://link.aps.org/doi/10.1103/PhysRevB.73.075107

[35] C. Vion, P. Spinicelli, L. Coolen, C. Schwob, J.-M. Frigerio, J.-P. Hermier, and A. Maître, "Controlled modification of single colloidal CdSe/ZnS nanocrystal fluorescence through interactions with a gold surface," *Opt. Express*, vol. 18, no. 7, pp. 7440–7455, Mar 2010. [Online]. Available: http://www.opticsexpress.org/abstract.cfm?URI=oe-18-7-7440

[36] R. Chance, A. Prock, and R. Silbey, "Comments on the classical theory of energy transfer," *The Journal of Chemical Physics*, vol. 62, no. 6, pp. 2245–2253, 1975.

[37] L. Novotny and B. Hecht, *Principles of Nano-Optics*, 2nd ed. Cambridge University Press, 2012.

[38] W. Lukosz, "Theory of optical-environment-dependent spontaneous-emission rates for emitters in thin layers," *Phys. Rev. B*, vol. 22, pp. 3030–3038, Sep 1980. [Online]. Available: https://link.aps.org/doi/10.1103/PhysRevB.22.3030

[39] D. Canneson, I. Mallek-Zouari, S. Buil, X. Quélin, C. Javaux, B. Dubertret, and J.-P. Hermier, "Enhancing the fluorescence of individual thick shell CdSe/CdS nanocrystals by coupling to gold structures," *New Journal of Physics*, vol. 14, no. 6, p. 063035, 2012. [Online]. Available: http://stacks.iop.org/1367-2630/14/i=6/a=063035

[40] H. Naiki, S. Masuo, S. Machida, and A. Itaya, "Single-photon emission behavior of isolated cdse/zns quantum dots interacting with the localized surface plasmon resonance of silver nanoparticles," *The Journal of Physical Chemistry C*, vol. 115, no. 47, pp. 23299–23304, 2011. [Online]. Available: https://doi.org/10.1021/jp207997j

[41] S. J. LeBlanc, M. R. McClanahan, M. Jones, and P. J. Moyer, "Enhancement of multiphoton emission from single cdse quantum dots coupled to gold films," *Nano Letters*, vol. 13, no. 4, pp. 1662–1669, 2013, pMID: 23510412. [Online]. Available: http://dx.doi.org/10.1021/nl400117h

[42] Y.-S. Park, Y. Ghosh, Y. Chen, A. Piryatinski, P. Xu, N. H. Mack, H.-L. Wang, V. I. Klimov, J. A. Hollingsworth, and H. Htoon, "Super-poissonian statistics of photon emission from single CdSe-

CdS core-shell nanocrystals coupled to metal nanostructures," *Phys. Rev. Lett.*, vol. 110, p. 117401, Mar 2013. [Online]. Available: https://link.aps.org/doi/10.1103/-PhysRevLett.110.117401

[43]   E. Ozbay, "Plasmonics: merging photonics and electronics at nanoscale dimensions," *science*, vol. 311, no. 5758, pp. 189–193, 2006.

[44]   F. Bigourdan, F. Marquier, J.-P. Hugonin, and J.-J. Greffet, "Design of highly efficient metallo-dielectric patch antennas for single-photon emission," *Opt. Express*, vol. 22, no. 3, pp. 2337–2347, Feb 2014. [Online]. Available: http://www.opticsexpress.org/abstract.cfm?URI=oe-22-3-2337

[45]   V. A. Markel and A. K. Sarychev, "Propagation of surface plasmons in ordered and disordered chains of metal nanospheres," *Phys. Rev. B*, vol. 75, p. 085426, Feb 2007. [Online]. Available: https://link.aps.org/doi/10.1103/PhysRevB.75.085426

[46]   M. Kuttge, E. Vesseur, J. Verhoeven, H. Lezec, H. Atwater, and A. Polman, "Loss mechanisms of surface plasmon polaritons on gold probed by cathodoluminescence imaging spectroscopy." *Applied Physics Letters*, vol. 93, no. 11, p. 113110, 2008.

[47]   P. Kambhampati, "Multiexcitons in semiconductor nanocrystals: A platform for optoelectronics at high carrier concentration," *The Journal of Physical Chemistry Letters*, vol. 3, no. 9, pp. 1182–1190, 2012. [Online]. Available: http://dx.doi.org/10.1021/jz300239j

[48]   J. S. Batchelder, A. H. Zewai, and T. Cole, "Luminescent solar concentrators. 1: Theory of operation and techniques for performance evaluation," *Appl. Opt.*, vol. 18, no. 18, pp. 3090–3110, Sep 1979. [Online]. Available: http://ao.osa.org/abstract.cfm?URI=ao-18-18-3090

[49]   L. R. Bradshaw, K. E. Knowles, S. McDowall, and D. R. Gamelin, "Nanocrystals for luminescent solar concentrators," *Nano Letters*, vol. 15, no. 2, pp. 1315–1323, 2015, pMID: 25585039. [Online]. Available: http://dx.doi.org/10.1021/nl504510t

[50]   F. Feng, L. T. Nguyen, M. Nasilowski, B. Nadal, B. Dubertret, L. Coolen, and A. Maître, "Consequence of shape elongation on emission asymmetry for colloidal CdSe/CdS nanoplatelets," *Nano Research*, vol. 11, no. 7, pp. 3593–3602, Jul 2018. [Online]. Available: https://doi.org/10.1007/s12274-017-1926-3

[51]   F. Feng, L. T. Nguyen, M. Nasilowski, B. Nadal, B. Dubertret, A. Maître, and L. Coolen, "Probing the fluorescence dipoles of single cubic CdSe/CdS nanoplatelets with vertical or horizontal orientations," *ACS Photonics*, vol. 5, no. 5, pp. 1994–1999, 2018. [Online]. Available: https://doi.org/10.1021/acsphotonics.7b01475

[52]    C. Lethiec, J. Laverdant, H. Vallon, C. Javaux, B. Dubertret, J.-M. Frigerio, C. Schwob, L. Coolen, and A. Maître, "Measurement of three-dimensional dipole orientation of a single fluorescent nanoemitter by emission polarization analysis," *Physical Review X*, vol. 4, no. 2, p. 021037, 2014.

**Acknowledgements**

The authors thank Willy Daney de Marcillac for spectrometric measurements, Loic Becerra, Bruno Gallas, Stéphane Chenot, Stéphan Suffit for precious advice and help in vapor deposition ellipsometry, and Jean-Jacques Greffet and François Marquier for fruitful discussions. This work was supported by the regional funding DIM NanoK through the project PATCH and by the ANR DELIGHT.

**Author contribution**

A.R.D., C.B., P.S., and A.M conceived and developed the fabrication protocol, A.R.D. and J.U.E performed the optical characterization, M.N. and B.D. synthesized the QDs, J.P.H. performed the simulations, A.R.D., A.M., C.S., L.C. and Z.W. investigated the optical characterization data, and A.R.D, A.M. and P.S. wrote the paper.

**Additional information**

The Supplementary Information details the fabrication of QDs and discusses their emission properties. The lithography protocol, and its lateral and vertical precision are explained in depth. The evidence of exciton and multiexciton emission in antenna emission is explored and validated through temporal filtering. Lastly, we elaborate on the estimation of the Purcell factor of the antenna of the paper.